\documentclass[preprint,12pt,TurnOnLineNumbers]{JASAnew}

\pdfoutput=1
\newcommand{\ek}{Simrad EK80}

\newcommand{\timesym}{t}
\newcommand{\freqsym}{f}
\newcommand{\samplesymt}{n}
\newcommand{\samplesymf}{m}
\newcommand{\genidxsym}{i}

\newcommand{\channelsym}{u}
\newcommand{\nchannels}{N_{\textrm{u}}}

\newcommand{\stagesym}{v}
\newcommand{\nstages}{N_{\textrm{v}}}

\newcommand{\fs}{f_{\textrm{s}}}
\newcommand{\fsdec}{f_{\textrm{s,dec}}}

\newcommand{\fc}{f_{\textrm{c}}}
\newcommand{\fn}{f_{\textrm{n}}}

\newcommand{\zrxe}{z_{\textrm{rx,e}}}
\newcommand{\ztde}{z_{\textrm{td,e}}}

\newcommand{\ptxe}{p_{\textrm{tx,e}}}
\newcommand{\prxe}{p_{\textrm{rx,e}}}

\newcommand{\tnom}{\tau}
\newcommand{\teff}{\tau_{\textrm{eff}}}

\newcommand{\ytxe}{y_{\textrm{tx,e}}}
\newcommand{\ytxa}{y_{\textrm{tx,a}}}
\newcommand{\yrxa}{y_{\textrm{rx,a}}}
\newcommand{\yrxe}{y_{\textrm{rx,e}}}

\newcommand{\ytx}{y_{\textrm{tx}}}
\newcommand{\ytxnorm}{\tilde{y}_{\textrm{tx}}}

\newcommand{\yrx}{y_{\textrm{rx}}}
\newcommand{\yrxorg}{y_{\textrm{rx,org}}}
\newcommand{\ymf}{y_{\textrm{mf}}}

\newcommand{\ypc}{y_{\textrm{pc}}}
\newcommand{\ypctarget}{y_{\textrm{pc,t}}}
\newcommand{\ypcspread}{y_{\textrm{pc,s}}}
\newcommand{\ymfauto}{y_{\textrm{mf,auto}}}
\newcommand{\ymfautored}{y_{\textrm{mf,auto,red}}}

\newcommand{\ptxauto}{p_{\textrm{tx,auto}}}

\newcommand{\ypctargetf}{Y_{\textrm{pc,t}}}
\newcommand{\ypctargetnormf}{\tilde{Y}_{\textrm{pc,t}}}
\newcommand{\ypcvolumef}{Y_{\textrm{pc,v}}}
\newcommand{\ypcvolumenormf}{\tilde{Y}_{\textrm{pc,v}}}

\newcommand{\ymfautof}{Y_{\textrm{mf,auto}}}
\newcommand{\ymfautoredf}{Y_{\textrm{mf,auto,red}}}

\newcommand{\prxetf}{P_{\textrm{rx,e,t}}}
\newcommand{\prxevf}{P_{\textrm{rx,e,v}}}

\newcommand{\decfac}{D}

\newcommand{\hfl}{h_{\textrm{fl}}}
\newcommand{\hannw}{w}
\newcommand{\hannwnorm}{\tilde{\hannw}}
\newcommand{\nw}{N_{\hannw}}

\newcommand{\tslide}{t_w} 

\newcommand{\bs}{\sigma_{\textrm{bs}}}
\newcommand{\mysp}{S_p}
\newcommand{\ts}{\textrm{TS}}
\newcommand{\sv}{S_{\textrm{v}}}

\newcommand{\range}{r}
\newcommand{\rangeref}{r_0}
\newcommand{\athw}{\phi}
\newcommand{\along}{\theta}
\newcommand{\gain}{g}
\newcommand{\gainzero}{g_0}
\newcommand{\eqang}{\psi}

\newcommand{\wlen}{\lambda}
\newcommand{\cw}{c}
\newcommand{\absorp}{\alpha}

\newcommand{\dft}{\textrm{DFT}}
\newcommand{\ndft}{{N_{\textrm{DFT}}}}

\newcommand{\atan}{\textrm{arctan2}}
\newcommand{\anglefalong}{\gamma_\along}
\newcommand{\anglefathw}{\gamma_\athw}

\begin{document}

\title[]{Quantitative processing of broadband data as implemented in a scientific splitbeam echosounder}

\author{Lars Nonboe Andersen}
\email{lars.nonboe.andersen@kongsberg.com}
\affiliation{Kongsberg Maritime AS, Strandpromenaden 50, 3191, Horten, Norway}

\author{Dezhang Chu}
\affiliation{Fishery Resource Analysis and Monitoring Division, Northwest Fisheries Science Center, National Marine Fisheries Service, National Oceanic and Atmospheric Administration, 2725 Montlake Blvd. E. Seattle, WA, 98112, USA}

\author{Harald Heimvoll}
\affiliation{Kongsberg Maritime AS, Strandpromenaden 50, 3191, Horten, Norway}

\author{Rolf Korneliussen}
\author{Gavin J. Macaulay}
\author{Egil Ona}
\affiliation{Ecosystem Acoustics, Institute of Marine Research, Bergen, 5001, Norway}

\begin{abstract}
The use of quantitative broadband echosounders for biological studies and surveys offers considerable advantages over narrowband echosounders. These include improved spectral-based target identification and significantly increased ability to resolve individual targets. Biological studies and surveys typically require accurate measures of backscatter strength and we present here a systematic and comprehensive explanation of how to derive quantitative estimates of target strength and volume backscattering, as a function of frequency from broadband echosounder signals.

\end{abstract}

\maketitle

\section{Introduction}

Management of fisheries resources typically requires knowledge of age and abundance structure over many years. For numerous fish species and stocks, abundance estimates can be obtained by spatially extensive surveys using quantitative echosounder systems, which measure the backscatter from fish \citep{Simmonds2005Fisheries}. An important property of scientific echosounders used for fisheries surveys is accurate measurement of the backscatter amplitude - this is of lesser importance for other echosounder applications, such as bathymetry, presence/absence of objects, and spatial properties of objects. Conventional echosounders generate an acoustic pulse with a narrow bandwidth (several kHz at most) and when several are operated simultaneously at widely spaced frequencies (such as 18, 38, 70, 120, 200, and 333 kHz) can help to categorize the backscatter into species or target categories. This is termed the multi-frequency approach \citep{Korneliussen2002operational, holliday1980}.

The application of acoustic pulses with a wide and continuous frequency range (broadband pulses) to fisheries applications is an obvious enhancement. Broadband pulses provide better frequency resolution and coverage than multi-frequency systems, i.e., a continuous frequency coverage over a wide frequency band, and with appropriate processing can provide significantly better along-beam resolution and a higher signal to noise ratio than narrowband pulses \citep{Chu1998Application, ehrenbergFMSlideChirp2000}. These benefits can lead to improved backscatter categorisation \citep{Traykovski1998Effect, Stanton2012Resonance, Korneliussen2016Acoustic, korneliussen2018} and hence more accurate abundance estimates.

There have been several scientific broadband echosounder systems developed for laboratory use \citep{Conti2003Wide-bandwidth, Forland2014Scattering, chu1992}, some prototype or custom-made systems \citep{Zakharia1989Wide-band, Zakharia1996Wideband, Simmonds1996Species, Foote2005Measuring, Imaizumi2009Detection, Briseno-Avena2015ZOOPS, Barr2002Target} and some commercially available systems \citep{Gordon1998FishMASS, Zedel2003Acoustic, Stanton2010New, ehrenbergFMSlideChirp2000, dennyBroadbandAcousticFish1998}. None of these systems have achieved widespread use in quantitative fisheries surveys presumably because the commercially-available systems have lacked one or more of the features that are very useful for an operational quantitative fisheries survey echosounder. These comprise (a) a large dynamic range receiver that does not saturate in normal conditions; (b) operating frequencies that are useful for detecting fish at several hundred metres range (between approximately 20 and 200 kHz); (c) split-aperture for detection of angle of arrival of echoes from individual organisms and calibration spheres; (d) capability to perform in-situ calibrations of the amplitude response; (e) a relatively easy transition from existing survey echosounders; (f) useful at typical ship survey speeds (e.g., 5 m s \textsuperscript{-1}). A recent addition to the set of commercially-available broadband echosounders is the Simrad EK80, which meets all of these features.

The purpose of this paper is to present a systematic and comprehensive explanation of how to derive quantitative broadband data from recorded broadband signals. Without loss of generality, we use the \ek\ as an example since it is currently the most commonly used broadband echosounder in the fisheries acoustics field. By presenting the design goals, implementation details, and recommended procedures and processing required to obtain quantitative broadband data, the authors hope to encourage and facilitate the realistic use of broadband signals in fisheries acoustics.

Our presentation uses nomenclature and approaches that are commonly used for narrowband echosounder systems, which were derived from radar processing \citep{cook1967}. In particular, the expressions for target strength ($\ts$) and volume backscattering strength ($\sv$) \citep{MacLennan2002consistent} are presented in a similar manner for broadband signals as for narrowband signals.

\section{Signal flow and initial processing}

\subsection{System overview}
\begin{figure}
\includegraphics[width=16cm]{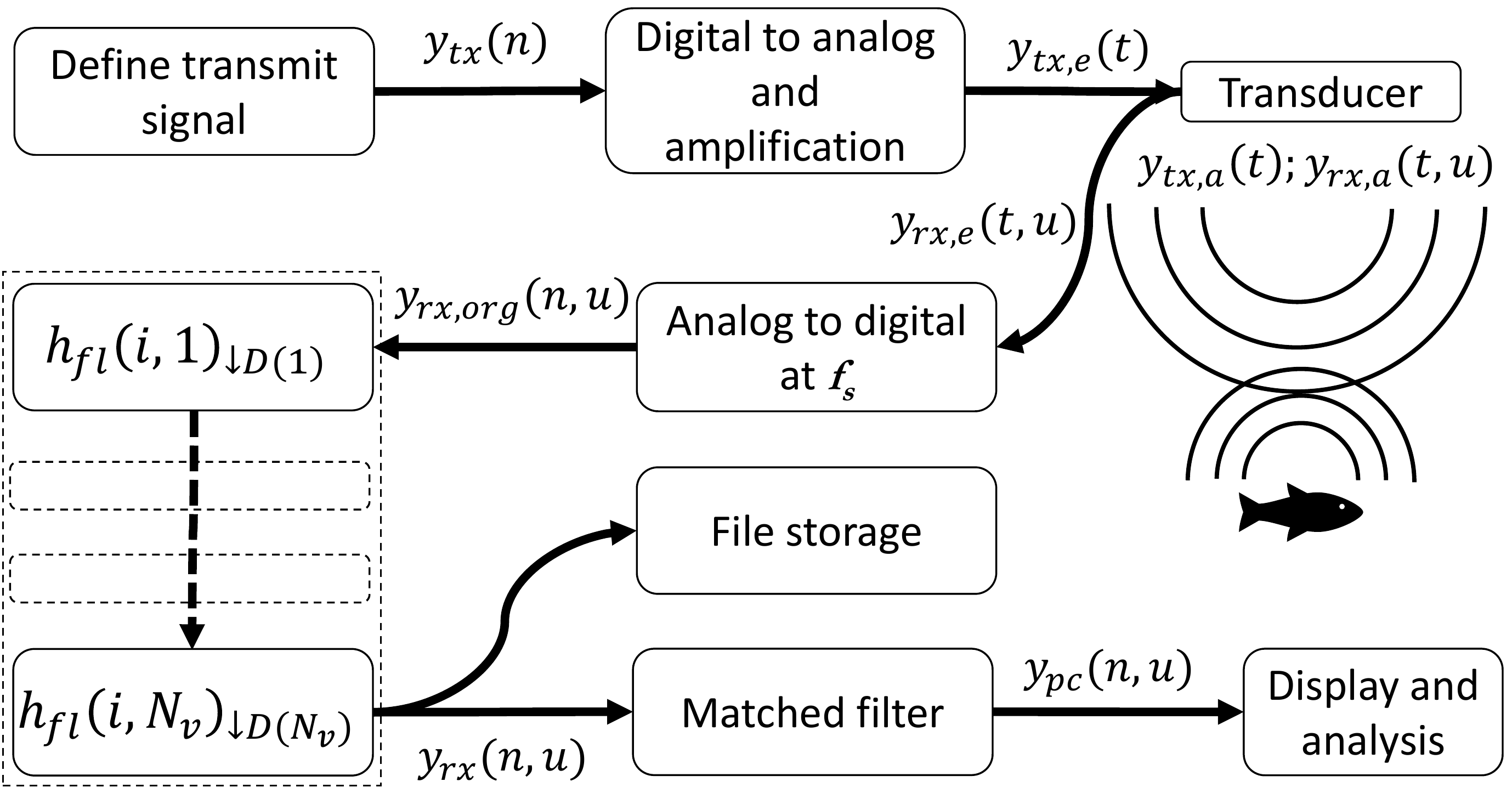}
\caption{\label{fi:ek_sys}Signal and data flow in the \ek{} system. An echosounder ping starts with the definition of a transmit signal (upper left) and ends with file storage (lower middle) and display and analysis (lower right). Note that all signals are complex-valued time-series.}
\end{figure}

A basic echosounder system consists of a transducer, a transceiver, and a computer program that controls the operation of the transceiver and records received signals. During transmission the program defines the signals which are created as electric signals in the transceiver, converted to acoustic signals by the transducer and transmitted into the water. The acoustic signals propagate through the water, are reflected or scattered by objects in the water, and propagate back to the transducer. During reception the transducer converts the received acoustic signals to electric signals, which are received, pre-amplified, filtered, digitized, and processed in the transceiver, and then transferred to the controlling program for further data processing and storage (Fig. \ref{fi:ek_sys}). Many types of transmit signals are feasible - this paper considers only the transmission of linear frequency modulated signals (also known as linear chirps).

\subsection{Internal signal processing}

The controlling computer program generates a short-duration digital transmit signal (a ping), $\ytx(\samplesymt)$, that is converted to an analogue electric signal, $\ytxe(\timesym)$, by the transceiver and applied to the transducer to generate the transmitted acoustic signal $\ytxa(\timesym)$ (Fig. \ref{fi:ek_sys}) where $\samplesymt$ is a sample index in the discrete time domain and $\timesym$ is time for the analog signal. Any returning acoustic signal, $\yrxa(\timesym)$, is received by each transducer sector, $\channelsym$, and converted to an analog electric signal, $\yrxe(\timesym,\channelsym)$, in the transducer and received by corresponding receiver channels, $\channelsym$, in the transceiver. For a split-aperture echosounder system, there are typically three or four channels in the system (a minimum of three are necessary for estimating the angle of arriving echoes). Here, we focus on a four-channel system.

The received electric signal, $\yrxe(\timesym,\channelsym)$, from each channel, $\channelsym$, is pre-amplified, filtered by an analog anti-aliasing filter, and digitized in the transceiver at a frequency of $\fs$, creating the digital signal, $\yrxorg(\samplesymt,\channelsym)$.

To remove noise and reduce the quantity of data, the sampled signal from each channel is filtered and decimated in multiple stages, $\stagesym$, using complex bandpass filters, $\hfl(\genidxsym,\stagesym)$, and decimation factors, $\decfac(\stagesym)$. The individual filter coefficients for each filter and decimation stage are indexed by $i$. The output signal from each channel, $\channelsym$, from each filter and decimation stage, $\stagesym$, is then given by:
\begin{equation}
\label{eq:yrx}
\yrx(\samplesymt,\channelsym,\stagesym) = \left( \yrx(\samplesymt,\channelsym,\stagesym-1) * \hfl(\genidxsym,\stagesym) \right)_{\downarrow \decfac(\stagesym)}, 
\stagesym = 1,\ldots,\nstages,
\end{equation}
where $\yrx(\samplesymt,\channelsym,0)$ is set to $\yrxorg(\samplesymt,\channelsym)$, being the signal before decimation, $*$ indicates convolution and $\nstages$ is the total number of filter stages. The output signal from the final filter and decimation stage, $\yrx(\samplesymt,\channelsym,\nstages)$, is shortened to $\yrx(\samplesymt,\channelsym)$ for convenience. For the output signal, $\yrx(\samplesymt,\channelsym)$, the decimated sampling rate, $\fsdec$, is given by:
\begin{equation}
\label{eq:fsdec}
\fsdec = \fs\prod_{\stagesym=1}^{\nstages} \frac{1}{\decfac(\stagesym)}.
\end{equation}
The characteristics of the bandpass filter and decimation factors are chosen with regard to the desired operating bandwidth, noise suppression levels, impulse response duration, and other common filter characteristics, with the aim of maintaining sufficient information in the data. The filtered and decimated complex samples from each transducer channel, $\yrx(\samplesymt,\channelsym)$, are considered raw data and are recorded together with additional data such as from position and motion sensors and system configuration data in raw data files for display and analysis by processing software.

\subsection{Pulse compression}
To increase signal-to-noise ratio and resolution along the acoustic beam a matched filter may be applied to the raw data samples \citep{turin1960}. This technique is also known as pulse compression \citep{klauder1960}. One approach for a matched filter is to use a normalized version of the ideal transmit signal as the replicate signal, filtered and decimated using the same filters and decimation factors as applied in Eq. \ref{eq:yrx}. The normalized ideal transmit signal, $\ytxnorm(\samplesymt)$, is given by:
\begin{equation}
\label{eq:ytxnorm}
\ytxnorm(\samplesymt) = \frac{\ytx(\samplesymt)}{\textrm{max}(\ytx(\samplesymt))}\end{equation}
where $\textrm{max}$ is the maximum value of $\ytx(\samplesymt)$. The filtered and decimated output signal, $\ytxnorm(\samplesymt,\stagesym)$, from each filter stage, $\stagesym$, using the normalized ideal transmit signal, $\ytxnorm(\samplesymt)$, as the input signal, is given by:
\begin{equation}
\label{eq:FilterStagesTX}
\ytxnorm(\samplesymt,\stagesym) = \left( \ytxnorm(\samplesymt,\stagesym-1) * \hfl(\genidxsym,\stagesym) \right)_{\downarrow \decfac(\stagesym)}, 
\stagesym = 1,\ldots,\nstages,
\end{equation}
where $\ytxnorm(\samplesymt,0)$ is set to $\ytxnorm(\samplesymt)$ and $\downarrow$ indicates decimation by the factor $\decfac(\stagesym)$. The output signal from the final filter and decimation stage, $\ytxnorm(\samplesymt,\nstages)$, is used as the matched filter and is denoted as $\ymf(\samplesymt)$.

\begin{figure}
\includegraphics{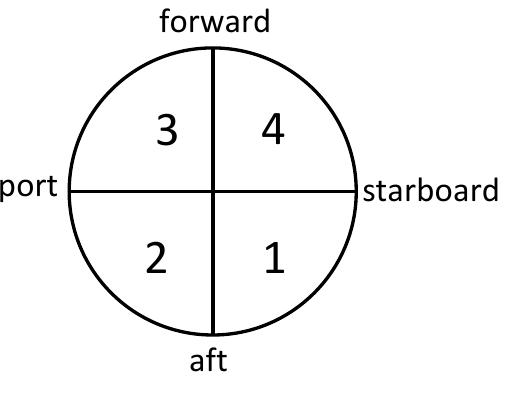}
\caption{\label{fi:trd_quad}Transducer divided into four quadrants. The labels are directions often used when a transducer is mounted on a ship.}
\end{figure}

To perform pulse compression the received signal, $\yrx(\samplesymt,\channelsym)$, is convolved with a complex conjugated and time-reversed version of the matched filter signal with the matched filter signal, and here also normalized with the $l^2$-norm of the matched filter to maintain received signal power. The pulse compressed signal, $\ypc(\samplesymt,\channelsym)$, then becomes
\begin{equation}
\label{eq:PulseComp}
\ypc(\samplesymt,\channelsym) = \frac{\yrx(\samplesymt,\channelsym)*\ymf^*(-\samplesymt)}{||\ymf||^2_2},
\end{equation}
where $||\ymf||$ indicates the $l^2$-norm of $\ymf$, also known as the Euclidean norm. The received power samples are then used to estimate target strength and volume backscattering strength. For estimating received power samples, the mean signal, $\ypc(\samplesymt)$, over all transducer sectors, $\nchannels$, will be used:
\begin{equation}
\label{eq:SumSig}
\ypc(\samplesymt) = \frac{1}{\nchannels} \sum_{\channelsym = 1}^{\nchannels} \ypc(\samplesymt,\channelsym).
\end{equation}
Compensation of echo strength for position in the acoustic beam requires an estimate of the echo arrival angle. This is obtained using the split-aperture method \citep{burdic1991}, which for broadband pules can be implemented with the angle values contained in the complex-valued $\ypc(\samplesymt)$ data, in combination with knowledge of transducer sector geometry. The principle is demonstrated with a transducer that is divided into four quadrants (Fig. \ref{fi:trd_quad}). In this example the summed signals from four halves (1+2, 2+3, 3+4, 4+1) are calculated as:
\begin{eqnarray}
\label{eq:SumHalves}
y_{\textrm{pc,fore}}(\samplesymt) & = & \frac{1}{2} \left( \ypc(\samplesymt,3)+\ypc(\samplesymt,4) \right),\\
y_{\textrm{pc,aft}}(\samplesymt) & = & \frac{1}{2} \left( \ypc(\samplesymt,1)+\ypc(\samplesymt,2) \right),\\
y_{\textrm{pc,star}}(\samplesymt) & = & \frac{1}{2} \left( \ypc(\samplesymt,1)+\ypc(\samplesymt,4) \right),\\
y_{\textrm{pc,port}}(\samplesymt) & = & \frac{1}{2} \left( \ypc(\samplesymt,2)+\ypc(\samplesymt,3) \right),
\end{eqnarray}
where fore, aft, star(board), and port indicate the relevant transducer halves.
\subsection{Auto correlation function}
The auto correlation function of the matched filter signal, $\ymfauto(\samplesymt)$, that will be used in a later processing step is defined as:
\begin{equation}
\label{eq:TXAuto}
\ymfauto(\samplesymt) = \frac{\ymf(\samplesymt)*\ymf^*(-\samplesymt)}{||\ymf||^2_2}.
\end{equation}

\subsection{Power and angle samples}
\begin{figure}
\includegraphics[width=\reprintcolumnwidth]{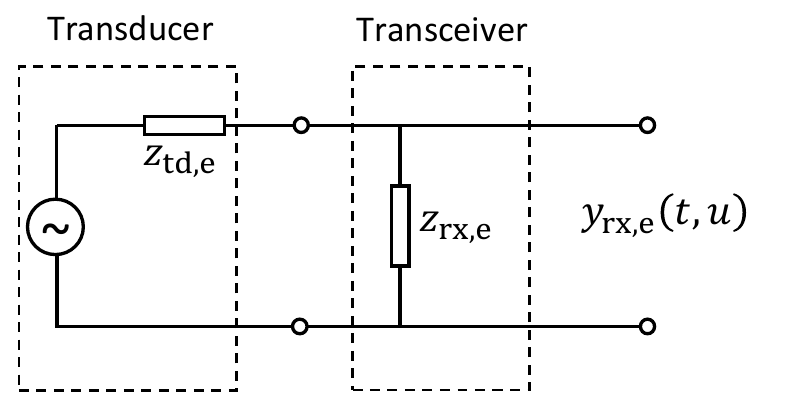}
\caption{\label{fi:impedances}Equivalent circuit diagram of transducer/transceiver with system impedances.}
\end{figure}

The transceiver measures voltage over a load, $\zrxe$, connected in series with the transducer impedance, $\ztde$. When calculating various acoustic properties a system gain parameter will be used which assumes a matched receiver load. The total received power, $\prxe(\samplesymt)$, from all transducer sectors for a matched receiver load (Fig. \ref{fi:impedances}) is given by: 
\begin{equation}
\label{eq:prx}
\prxe(\samplesymt) = \nchannels\left( \frac{|\ypc(\samplesymt)|}{2\sqrt{2}} \right)^2 \left( \frac{|\zrxe+\ztde|}{\zrxe} \right)^2 \frac{1}{|\ztde|}.
\end{equation}

Forward/aft and port/starboard phase angles of target echoes are estimated by combining the transducer half signals thus: 
\begin{eqnarray}
\label{eq:phase1}
y_\along(\samplesymt) & = & y_{\textrm{pc,fore}}(\samplesymt) y_{\textrm{pc,aft}}^*(\samplesymt), \\
y_\athw(\samplesymt) & = & y_{\textrm{pc,star}}(\samplesymt) y_{\textrm{pc,port}}^*(\samplesymt),
\end{eqnarray}
where $y_\along(\samplesymt)$ is the electrical angle along the minor axis of the transducer (positive in the forward direction when ship-mounted) and $y_\athw(\samplesymt)$ the electrical angle along the major axis of the transducer (positive to starboard when ship-mounted), where complex signals are represented in the form $e^{j 2\pi \freqsym \timesym}$, where $j = \sqrt{-1}$. The physical echo arrival angles ($\along$ and $\athw$) are then given by:
\begin{eqnarray}
\label{eq:phase2}
\along(\samplesymt) & = & \arcsin\left( \frac{\atan\left( \Im(y_\along(\samplesymt)), \Re(y_\along(\samplesymt) \right)}{\anglefalong}\right) \\
\athw(\samplesymt) & = & \arcsin\left( \frac{\atan\left( \Im(y_\athw(\samplesymt)), \Re(y_\athw(\samplesymt) \right)}{\anglefathw}\right).
\end{eqnarray}
where $\anglefalong$ and $\anglefathw$ are constants that convert from phase angles to physical echo arrival angles and are derived from the transducer geometry and $\fc$ the centre frequency of the chirp pulse \citep{ehrenberg1979}. The inverse sine is indicated by $\arcsin$, the four quadrant inverse tangent which returns values in the interval $[-\pi, \pi]$ inclusive is indicated by $\atan$, the real part of a complex number by $\Re$ and the imaginary part by $\Im$. As a mnemonic, the horizontal line in the symbol used for the forward/aft direction, $\along$, represents the pivot axis for the alongship angles and the near-vertical line in the $\athw$ symbol indicates the pivot axis for port/starboard angles.

\section{Target strength}

Echoes from single targets are often characterised by their $\ts$, which is related to the differential backscattering cross section, $\bs$, via
\begin{equation}
\label{eq:TS_bs}
\ts = 10\log_{10}\left(\frac{\bs}{\rangeref^2}\right),
\end{equation}
where $\log_{10}$ is the logarithm with base 10 and $\rangeref$ is 1 m.

The power-budget equation (i.e., sonar equation) for a single target \citep[Formulation D, ][]{lunde2016} at frequency $\freqsym$ is:
\begin{equation}
\label{eq:TS}
\ts(\freqsym) = 10\log_{10}(\prxetf(\freqsym)) + 40\log_{10}(\range) + 2\absorp(\freqsym) \range 
- 10\log_{10}\left( \frac{\ptxe \wlen^2 \gain^2(\along,\athw,\freqsym)}{16\pi^2} \right),
\end{equation}
where $\prxetf(\freqsym)$ is the Fourier transform of the received electric power in a matched load for a signal from a single target at frequency $\freqsym$, $\range$ is the range to the target, $\absorp$ the acoustic absorption at frequency $f$, $\ptxe$ the transmitted electric power, $\wlen$ the acoustic wavelength, and $\gain$ the transducer gain along the main acoustic axis, incorporating beam compensation based on the estimated target bearing, $(\along,\athw)$.

The point scattering strength, $\mysp(\samplesymt)$, is estimated by applying Eq. \ref{eq:TS} to the received digitized power samples using the on-axis gain value with $f$ set to the centre frequency of the broadband pulse, $\fc$: 
\begin{equation}
\label{eq:Sp}
\mysp(\samplesymt) = 10\log_{10}(\prxe(\samplesymt)) + 40\log_{10}(\range(\samplesymt)) 
+ 2\absorp(\fc) \range(\samplesymt) - 10\log_{10}\left( \frac{\ptxe \wlen^2(\fc) \gainzero^2(\fc)}{16\pi^2} \right),
\end{equation}
noting that $\mysp(\samplesymt)$ is an average over frequency of all echoes from single or multiple targets received at sample $\samplesymt$.

Based on the point scattering strength samples and the phase angle samples, single targets can be detected, and range and bearing to the single targets can be estimated.

From the pulse compressed data, $\ypc(\samplesymt)$, the samples before and after the peak echo level corresponding to an echo from a single target are extracted to obtain the signal, $\ypctarget(\samplesymt)$, noting that the number of samples before the peak can differ from those after the peak. From the autocorrelation function of the matched filter signal, $\ymfauto(\samplesymt)$, the equivalent samples around the peak are extracted to create the reduced autocorrelation signal of the matched filter signal, $\ymfautored(\samplesymt)$. Depending on the target scattering characteristics and the distance to any adjacent single targets, the number of samples around the peak echo level in $\ypctarget(\samplesymt)$ that contain the majority of the echo energy can be more or less than the total number of samples around the peak of $\ymfauto(\samplesymt)$. If the number of samples around the target is more than the total number of samples around the peak of $\ymfauto(\samplesymt)$ all samples around the peak of $\ymfauto(\samplesymt)$ are used. If the number of samples around the target is less than the total number of samples around the peak of $\ymfauto(\samplesymt)$, this lower number is used to create a reduced autocorrelation signal, $\ymfautored(\samplesymt)$.

The discrete Fourier transforms of the target signal, $\ypctargetf(\samplesymf)$, and the reduced auto correlation signal, $\ymfautoredf(\samplesymf)$, are given by:
\begin{eqnarray}
\label{eq:DFT_Target_Auto}
\ypctargetf(\samplesymf) & = & \dft_\ndft(\ypctarget(\samplesymt)),\\
\ymfautoredf(\samplesymf) & = & \dft_\ndft(\ymfautored(\samplesymt)),
\end{eqnarray}
where $\dft$ indicates the Fourier transform of length $\ndft$ and $\samplesymf$ the sample index in the frequency domain.
The nomalized discrete Fourier transform of the target signal, $\ypctargetnormf(\samplesymf)$, is then calculated by: 
\begin{equation}
\label{eq:DFT_Target_Auto_Norm}
\ypctargetnormf(\samplesymf) = \frac{\ypctargetf(\samplesymf)} {\ymfautoredf(\samplesymf)}.
\end{equation}
Assuming, as a first approximation, that the impedances of the transceiver and transducer are independent of frequency, the received power into a matched load, $\prxetf(\samplesymf)$, is then estimated by:
\begin{equation}
\label{eq:prx_FFT_target}
\prxetf(\samplesymf) = \nchannels\left( \frac{|\ypctargetnormf(\samplesymf)|}{2\sqrt{2}} \right)^2 
\left( \frac{|\zrxe+\ztde|}{|\zrxe|}\right)^2 \frac{1}{|\ztde|}, 
\end{equation}
noting that any variation of impedance with frequency will be reflected in the $\gainzero$ obtained from the calibration process.

Target strength can then be estimated using Eq. \ref{eq:TS}, but with $\freqsym$ replaced by the discrete index of frequency, $\samplesymf$:
\begin{equation}
\label{eq:TS_f}
\ts(\samplesymf) = 10\log_{10}(\prxetf(\samplesymf)) + 40\log_{10}(\range) + 2\absorp(\samplesymf)\range 
- 10\log_{10}\left( \frac{\ptxe \wlen_\samplesymf^2 \gain^2(\along,\athw,\samplesymf)}{16\pi^2} \right).
\end{equation}

\section{Volume backscattering strength}

Echoes from multiple scatterers can be quantified using volume backscattering strength, $\sv$, being the density of backscattering cross sections, and is given by:
\begin{equation}
\label{eq:sv}
\sv  =  10\log_{10}\frac{\sum\bs}{V}.
\end{equation}
where $V$ is the volume occupied by the scattering targets. The power-budget equation for multiple targets is then:
\begin{equation}
\label{eq:sv_f}
\sv(\freqsym) = 10\log_{10}(\prxevf(\freqsym)) + 20\log_{10}(\range) + 2\absorp(\freqsym)\range 
- 10\log_{10}\left( \frac{\ptxe \wlen^2 \cw \tslide \eqang(\freqsym) \gainzero^2(\freqsym)}{32\pi^2} \right), 
\end{equation}
where $\prxevf(\freqsym)$ is the received electric power in a matched load for the signal from a volume at frequency $\freqsym$, $\cw$ the sound speed, and $\tslide$ the duration of the time window, excluding the zero-padded portion if applied, used for evaluating the frequency spectrum. Note that $\range$ is the range to the centre of the range volume covered by $\tslide$, and the two-way equivalent beam angle, $\eqang$, is a function of frequency that is derived from an empirical estimate of $\eqang$ at the nominal frequency, $\fn$:
\begin{equation}
\label{eq:PsiFc}
\eqang(f) = \eqang(\fn)\left(\frac{\fn}{f}\right)^2.
\end{equation}

Volume backscattering samples compressed over the operational frequency band are estimated by applying Eq. \ref{eq:sv_f} to the received digitized power samples using the on-axis gain value with $f$ set to the centre frequency of the broadband pulse, $\fc$:
\begin{equation}
\label{eq:Sv}
\sv(\samplesymt)  =  10\log_{10}(\prxe(\samplesymt)) + 20\log_{10}(\range(\samplesymt)) + 2\absorp(\fc)\range(\samplesymt) 
- 10\log_{10}\left( \frac{\ptxe \wlen^2(\fc) \cw \teff \eqang(\fc) \gainzero^2(\fc)}{32\pi^2} \right),
\end{equation}
noting that $\sv(\samplesymt)$ is an average over frequency of all echoes received at sample $\samplesymt$. In this case, the time window, $\tslide$, is the effective pulse duration, $\teff$, resulting from pulse compression. The effective pulse duration is defined as the pulse duration at transmit power $\ptxe$ which produces the same energy as the actual transmitted pulse, and is estimated from digitized data via:
\begin{eqnarray}
\label{eq:TauEff}
\ptxauto(\samplesymt) & = & |\ymfauto(\samplesymt)|^2,\\
\teff & = & \frac{\sum \ptxauto(\samplesymt)}{\textrm{max}(\ptxauto(\samplesymt))\fsdec},
\end{eqnarray}
where $\ptxauto$ is the square of the absolute value of the matched filter autocorrelation function, and the summation is calculated over a duration of twice the nominal pulse duration, $2\tnom$. For an ideal system, i.e., no tapering at the rising and trailing edges of the transmitted signal, the effective pulse duration is the same as the transmit pulse duration.

To estimate $\sv$ as a function of frequency a Fourier transform is used, repeatedly applied via a sliding window in range. However, the duration of this sliding window can be so long that the difference in spherical spreading loss compensation ($r^2$, implemented as the $20\log_{10}(\range)$ term in Eq. \ref{eq:Sv}) from the beginning of the window to the end can be significant, particularly for short range measurements. Thus, compensation for spreading loss is performed before applying the discrete Fourier transform. Absorption loss compensation is also range dependent (and frequency dependent), but since absorption loss compensation is insignificant for typical operating frequencies at short ranges and the difference in absorption loss compensation between the beginning and the end of the sliding window is insignificant at longer ranges, compensation for absorption loss is performed after applying the discrete Fourier transform.

Compensation of spherical spreading loss requires compensation of received power by a factor of $r^2$, and hence compensation of amplitude by a factor of $\range$:
\begin{equation}
\label{eq:spreadcomp}
\ypcspread(\samplesymt) = \ypc(\samplesymt)\range(\samplesymt).
\end{equation}
where $\ypcspread(\samplesymt)$ is the pulse compressed signal compensated for spherical spreading. A discrete Fourier transform is performed on the range compensated pulse compressed sample data using a normalized sliding Hanning window, $\hannw(\genidxsym)$. The duration, $\tslide$, of the sliding window is chosen as a compromise between along-beam range resolution and frequency resolution. We suggest that it be at least twice the pulse duration and for computational efficiency reasons should result in a number of samples, $\nw$, which is a power of 2.

The normalised Hanning window, $\hannwnorm$, is given by: 
\begin{equation}
\label{eq:hannw}
\hannwnorm(\genidxsym) = \frac{\hannw(\genidxsym)}{\left( \frac{||\hannw||_2}{\sqrt{\nw}} \right)}, i = \frac{-\nw}{2}, \ldots, \frac{\nw}{2}
\end{equation}
and the discrete Fourier transform of the windowed data, $\ypcvolumef(\samplesymf)$, is then obtained from:
\begin{equation}
\label{eq:FFT_volume}
\ypcvolumef(\samplesymf) = \dft_\ndft 
\left( \hannwnorm(\genidxsym) \left(\ypcspread (\genidxsym+\samplesymt) \left[ u(\genidxsym + \frac{\nw}{2}) - u(\genidxsym - \frac{\nw}{2}) \right] \right) \right),
\end{equation}
where $u(\genidxsym)$ is the step function and $\samplesymt$ is the sample data index for the centre of the sliding window. The discrete Fourier transform of the auto correlation function of the matched filter signal, $\ymfautof(\samplesymf)$, also needs to be evaluated at the same frequencies:
\begin{equation}
\label{eq:FFT_TX_Auto}
\ymfautof(\samplesymf)  =  \dft_\ndft (\ymfauto(\samplesymt)).
\end{equation}
The normalized discrete Fourier transform of the windowed data, $\ypcvolumenormf(\samplesymf)$, is then given by:
\begin{equation}
\label{eq:FFT_volume_norm}
\ypcvolumenormf(\samplesymf) = \frac{\ypcvolumef(\samplesymf)}{\ymfautof(\samplesymf)},
\end{equation}
and received power into a matched load, $\prxevf(\samplesymf)$, is estimated from:
\begin{equation}
\label{eq:prx_FFT_volume}
\prxevf(\samplesymf) = \nchannels \left( \frac{|\ypcvolumenormf(\samplesymf)|}{2\sqrt{2}} \right)^2 \left( \frac{|\zrxe+\ztde|}{|\zrxe|}\right)^2 \\
\frac{1}{|\ztde|}. 
\end{equation}
Finally, the discretized estimate of $\sv(\freqsym)$, $\sv(\samplesymf)$, is given by:
\begin{equation}
\label{eq:Sv_FFT}
\sv(\samplesymf) = 10\log_{10}(\prxevf(\samplesymf)) + 2\absorp(\samplesymf) \range \\
- 10\log_{10}\left( \frac{\ptxe \wlen^2 \cw \tslide \eqang(\samplesymf) \gainzero^2(\samplesymf) }{32\pi^2} \right).
\end{equation}

\section{Illustrative examples}

\begin{figure}
\includegraphics[width=16cm]{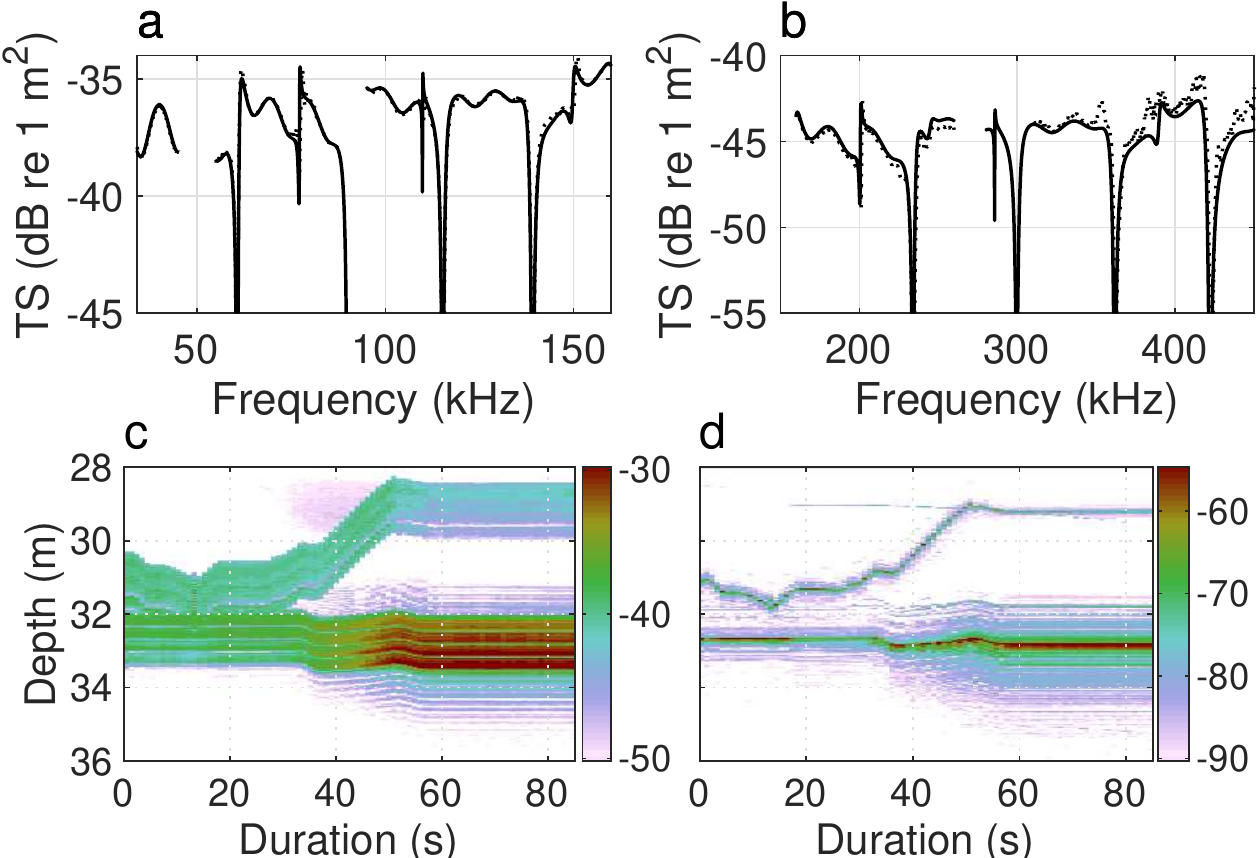}
\caption{\label{fi:ts_example}(color online) Target strength of 57.2 mm (a) and 22.0 mm (b) diameter tungsten carbide (with 6\% cobalt binder) calibration spheres, theoretical (solid line) and measured and derived using Eq. (\ref{eq:TS_f}) (dotted line). Echogram of non pulse compressed backscatter [dB re 1 W] (c) and pulse compressed backscatter [dB re 1 $\textrm{m}^2$] (d) from 57.2 mm (lower band) and 22.0 mm (upper band) diameter tungsten carbide spheres using a 2.048 ms duration linear frequency modulated pulse (160--260 kHz). The range from the spheres to the transducer varied during the time period shown.}
\end{figure}

A frequency modulated pulse scattered by a metallic sphere will exhibit frequencies at which very little energy is returned due to destructive interference \citep{stanton2008}. This is visible in the $\ts$ (Fig. \ref{fi:ts_example}a, \ref{fi:ts_example}b) and agrees well with theoretical estimates of the backscatter from spheres \citep{maclennan1981}. The amplitude of the backscatter signal also clearly shows these nulls (Fig. \ref{fi:ts_example}c), which are readily visible here due to the use of a linear chirp where time through the pulse corresponds to specific frequencies. The marked increase in range resolution is apparent once pulse compression has been applied (Fig. \ref{fi:ts_example}d), as are the temporal effects of the pulse compression operation.

A metallic sphere is a rather simple and ideal scatterer and we also present $\sv$ from a school of Atlantic mackerel (\textit{Scomber scombrus}) (Fig. \ref{fi:sv_example}a). The trend for increasing $\sv$ with frequency is well-known \citep{korneliussen2010} and is consistent with the trend observed in this example. In contrast to data from isolated scatterers, such as metallic spheres, the benefit of pulse compression on the backscatter from an object that generates many overlapping echoes is not immediately obvious (Fig. \ref{fi:sv_example}b, \ref{fi:sv_example}c), although in regions where the fish density decreases (e.g., top left of the school), single target echoes become visible.

\begin{figure}
\includegraphics[width=16cm]{Figure5}
\caption{\label{fi:sv_example}(color online) (a): Mean frequency response (solid line) and standard error (dashed line) from a school of Atlantic mackerel (\textit{Scomber scombrus}) from three simultaneously operating linear frequency modulated echosounder channels (90--170, 160--260, and 280--450 kHz). Echogram of backscattered power [dB re 1 W] (b) and pulse compressed $\sv$ [dB re 1 $m^{-1}$] (c) from the 160--260 kHz pulse. The school is the trapezoid-shaped object between 25 and 90 m below the surface. The echo from the seafloor is visible at 130 m depth.} \end{figure}

\section{Discussion}
Obtaining quantitative broadband data is more complicated than for narrowband echosounders, due in part to the need to account for frequency dependence in most variables and parameters, and the need for an increased understanding of digital signal processing techniques. For example, the \ek{} currently provides complex demodulated \citep{hasan1983} voltages for each transducer sector rather than derived quantities (e.g., the envelope of a pulse compressed signal and the frequency response of single targets and volumes) - these outputs must instead be calculated as proposed in this paper. This was an implementation decision made to allow for more flexible use of the echosounder data, and to ease the development of new processing methodologies. This decision has several disadvantages, such as the significantly increased data quantity and markedly higher amount of computation required to simply display an echogram. These can be ameliorated to some degree by processing the data into a more directly useful form before storage. The advantages of having access to unprocessed data were considered to out-weigh these disadvantages due to the potential benefits of more sophisticated uses of the acoustic data, especially for a tool that is only beginning to be applied to the field of fisheries acosutics.

The methodology to process broadband data has been presented in a general form in the previous sections without any accord to engineering limits. However, any physical implementation introduces operational constraints. As an example, a system designed for shipboard installations with ample access to electrical and computer processing power and large data storage will typically have different operational contraints compared to a system intended for autonomous platforms where electrical power and computing resources can be severely limited. In addition, the transducer is a significant constraint on the operational parameters of an echosounder and is usually the main determinant of the usable transceiver operating parameters (such as transmit bandwidth, maximum transmit power, pulse duration, ping rate, etc).

The use of broadband signals in fisheries acoustics is a developing area and we anticipate many valuable enhancements will occur in the coming years. For example, the use of $\ts(\freqsym)$ and $\sv(\freqsym)$ to improve acoustic target classification \citep{bassettBroadbandEchosounderMeasurements2018, korneliussen2018}, and the potential of the high range resolution from pulse compression to observe small-scale fish behviours \citep{skaret2020} and to detect objects adjacent to boundaries \citep{lavery2017}. The basic formulation for calculating $\ts(\freqsym)$ and $\sv(\freqsym)$ presented here provides the foundation for future enhancements.

The formulation presented in this paper results in several frequency dependent parameters, such as transducer gain, two-way equivalent beam angle, and the water absorption coefficient, that are required to quantitatively estimate $\ts(\freqsym)$ and $\sv(\freqsym)$ from received broadband signals. Methods to estimate these are not within the scope of this paper, but common practise is to use the conventional sphere backscatter calibration methodology \citep{demerCalibrationAcousticInstruments2015} slightly enhanced for broadband \citep{hobaekCharacterizationTargetSpheres2013, lavery2017}. We note that these methods do not provide an operational method to estimate $\teff$ or $\eqang(\freqsym)$, especially for ship-mounted transducers, and that empirical measurements of these parameters are necessary to fully calibrate both narrowband and broadband echosounders.

The processing equations and methodology presented in this paper have been implemented in version 1.12.4 and earlier of the \ek{} software.

\section{Conclusion}

A set of equations for calculating calibrated, frequency-dependent, target strength and volume backscatter from broadband echosounder signals have been presented, with reference to the \ek{} echosounder.

\section{Data Availability Statement}

The data associated with this article are available on request from the authors.


\end{document}